\documentclass[prd,aps,showpacs,,nofootinbib,superscriptaddress,preprintnumbers,epsf,psf]{revtex4}
\usepackage[english]{babel}
\usepackage{pstricks}
\usepackage[dvips]{graphicx}
\usepackage{epsf}
\usepackage{amsmath} 
\usepackage{amssymb}

\begin{document}

\newcommand{\zbar}{\bar{z}}
\newcommand{\odd}{\mathbb{O}}
\newcommand{\pom}{\mathbb{P}}
\newcommand{\tmin}{t_{\rm min}}
\newcommand{\smin}{s_{\rm min}}
\newcommand{\gev}{{\rm GeV}}

\preprint{\begin{tabular}{l}
      CPHT-RR078.1008 \\
      LPT-ORSAY 08/84\\
~
    \end{tabular}
 }

\title{Hard Pomeron-Odderon interference effects in the production of $\pi^{+}\pi^{-}$
pairs in high energy $\gamma\gamma$ collisions at the LHC}

\author{ B. Pire}
\address{CPhT, {\'E}cole Polytechnique, CNRS, 91128 Palaiseau, France }
\author{ F. Schwennsen}
\address{CPhT, {\'E}cole Polytechnique, CNRS, 91128 Palaiseau, France }
\address{ LPT, Universit{\'e} Paris-Sud, CNRS, 91405 Orsay, France}
\author{L. Szymanowski}
\address{Soltan Institute for Nuclear Studies, Warsaw, Poland}
\author{S. Wallon}
\address{ LPT, Universit{\'e} Paris-Sud, CNRS, 91405 Orsay, France}

\begin{abstract}
We estimate the production of two meson pairs in high energy photon photon collisions produced
in ultraperipheral collisions at LHC. We show that the study of charge asymmetries may reveal the existence of the perturbative Odderon and discuss the concrete event rates expected at the LHC. Sizable  rates and asymmetries are expected  in the case of proton-proton  collisions and medium values of $\gamma \gamma$ energies $\sqrt{s_{\gamma \gamma}} \approx 20\,{\rm GeV}$. 
Proton-proton collisions will benefit from a high rate due to a large effective $\gamma\gamma$ luminosity and ion-ion collisions with a somewhat lower rate from the possibility to trigger on ultraperipheral collisions and a reduced background from strong interactions.
\end{abstract}


\maketitle
\narrowtext

\noindent
\section{ Introduction}

Hadronic reactions at low momentum transfer and high energies are described
 in the framework of QCD in terms of the dominance of color singlet
exchanges corresponding to a few reggeized gluons. 
The charge conjugation even sector of the $t-$channel exchanges is
understood as the QCD-Pomeron. The charge-odd exchange is less well understood, although the 
need for the Odderon contribution \cite{LN,Doschrecent}, in particular to understand the different
behaviors of $pp$ and $\bar p p$ elastic cross sections \cite{Breakstone:1985pe}, is quite generally accepted.
Studies of specific channels where the Odderon contribution is expected
to be singled out have turned out to be very disappointing \cite{Olsson:2001nm} but new channels have recently
been proposed \cite{Bzdak:2007cz,Szymanowski:2007qb,Szymanowski:2007bv,Braunewell:2004pf}.  
Another  strategy to reveal the Odderon  has been first initiated
in Ref. \cite{Brodsky:1999mz}, where it was stressed that the study of observables where Odderon effects
are present  at the amplitude level -- and not at the squared amplitude
level -- is mandatory to get a convenient sensitivity to a rather small normalization
of this contribution. This approach has been extended to the production of pion pairs
 \cite{Ginzburg:2002zd,Ginzburg:2003ci,Hagler:2002nh,Hagler:2002sg,Hagler:2002nf,Hagler:2002cq} since the $\pi^+\,\pi^-$-state
does not have  any definite charge parity and therefore both Pomeron and
Odderon exchanges may contribute to the production amplitude. 

\begin{figure}[t]
\centerline{\includegraphics[scale=.5]{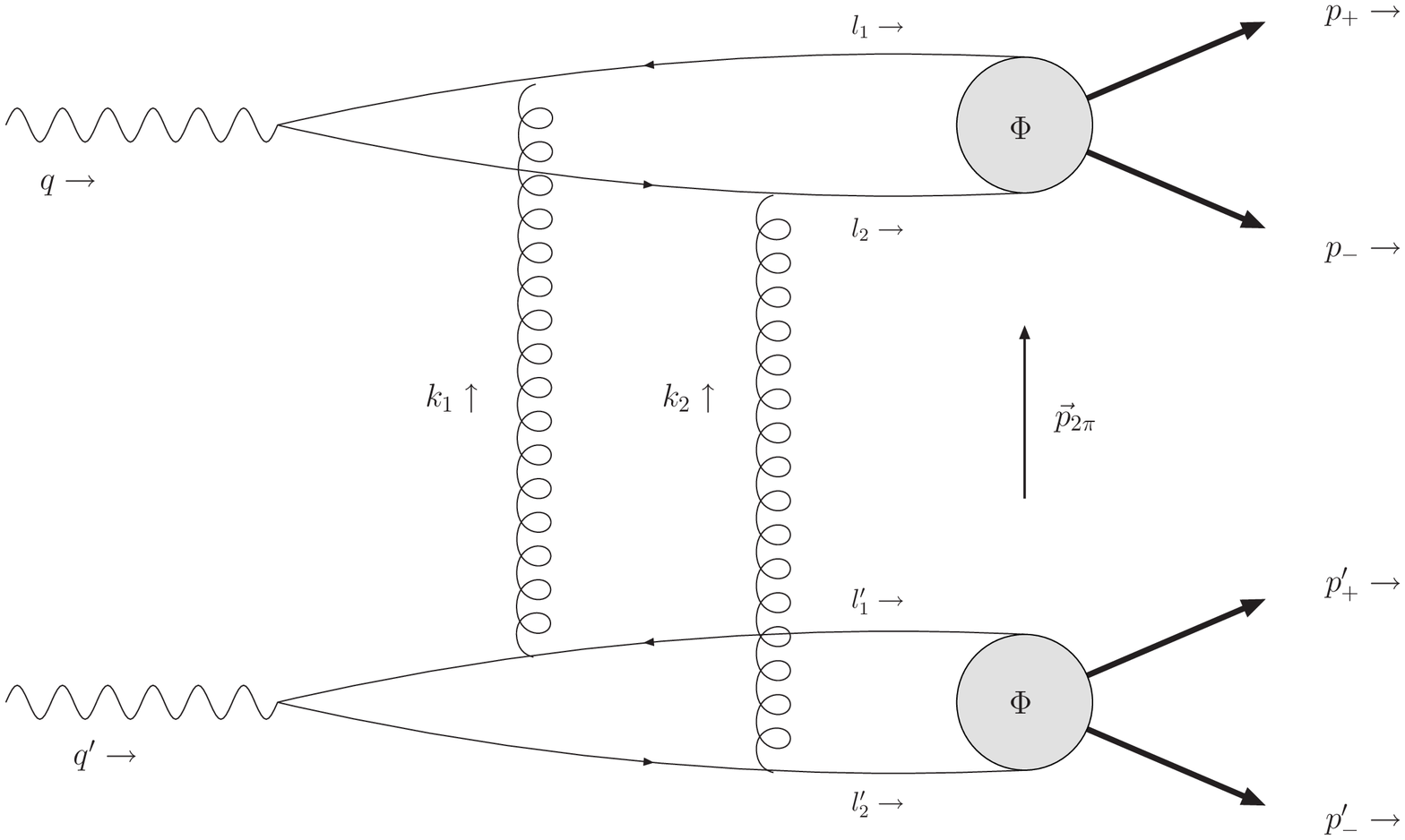}}
\caption{{\protect\small Kinematics of the reaction $\gamma \gamma \to \pi^+ \pi^-\;\; \pi^+ \pi^-$ in a sample Feynman diagram of the two gluon exchange process.}}
\label{fig:1}
\end{figure}

In this paper, we study within perturbative QCD (pQCD)  the
charge  asymmetries in the  production of two pion pairs in photon-photon collisions
\begin{equation}  
\gamma (q, \varepsilon)\;\; \gamma (q', \varepsilon') \to \pi^+(p_+)\;\; \pi^-(p_-)\;\;
\pi^+(p'_+)\;\; \pi^-(p'_-)\;, \label{gg}
\end{equation}
where $\varepsilon$ and $\varepsilon'$ are the initial photon polarization vectors, see Fig.~\ref{fig:1}. 
We have in mind the ultraperipheral collisions of protons or nuclei of high energies, which constitute a promising new way to 
study QCD processes initiated by two quasi real photons \cite{Baur:2001jj,Baltz:2007kq,Schicker:2008ih}. 
At high energy the application of pQCD for the calculation of a part of this process is justified by the presence of a 
 hard scale: the momentum transfer $t = (q - p_+ -p_-)^2$ from an initial photon to a final pion pair,
$-t$ being of the order of a few GeV$^2$.
The amplitude of this process may be calculated within $k_T$-factorization, as the convolution of two impact factors representing the 
photon to pion pair transitions and a two or three gluon exchange. The impact factors themselves include the convolution of a
perturbatively calculable hard part with a non-perturbative
input, the two pion generalized distribution amplitudes (GDA) which parametrize the quark-antiquark to hadron transition. Since
the $\pi^+\pi^-$ system is not a  charge parity
eigenstate, the GDA includes two charge parity components and allows for
a study of the corresponding interference term. The relevant GDA is here
just given by the light cone wave function of the two pion system \cite{Diehl:2000uv}.
 
\section{Kinematics}

Let us first specify the kinematics of the process under study, namely the
photon-photon scattering as stated in Eq.~(\ref{gg}).
We introduce a Sudakov representation of all particle momenta using the Sudakov
light-like momenta $p_1,p_2$. The initial photon momenta are
written as
\begin{equation}
q^{\mu }=p_{1}^{\mu },\hspace{1cm} q'^{\mu }=p_{2}^{\mu },
\label{q}
\end{equation}
where $s = 2p_1 \cdot p_2$. 
The momenta of the two pion systems are given by
\begin{eqnarray}  \label{2pi}
p_{2\pi}^{\mu} &=&   (1-\frac{\vec{p}_{2\pi}^{\;2}}{s})p_1^{\mu} 
                   + \frac{m_{2\pi}^2 + \vec{p}_{2\pi}^{\;2} }{s}p_2^{\mu} 
                   + p_{2\pi \perp}^{\mu},
               \quad p_{2\pi\perp}^2=-\vec{p}_{2\pi}^{\;2}, \nonumber\\
p_{2\pi}^{'\mu} &=& (1-\frac{\vec{p}_{2\pi}^{\;'2}}{s})p_2^{\mu} 
                   +  \frac{m_{2\pi}^{'2} + \vec{p}_{2\pi}^{\;'2} }{s}p_1^{\mu} 
                   + p_{2\pi \perp}^{'\mu},
               \quad p_{2\pi\perp}^{'2}=-\vec{p}_{2\pi}^{\;'2}  .
\end{eqnarray}

The (massless) quark momentum $l_1$ and antiquark momentum $l_2$ inside the upper loop before
the formation of  the two pion system are parametrized as
\begin{equation}  \label{l1}
l_1^{\mu} = z p_1^{\mu} 
     + \frac{(\vec{l}+z\,\vec{p}_{2\pi})^2}{zs}p_2^{\mu}
     + (l_\perp +z\, p_{2\pi \perp})^{\mu} ,
\end{equation}
\begin{equation}  \label{l2}
l_2^{\mu} = \bar{z} p_1^{\mu} 
     + \frac{(-\vec{l}+ \bar{z}\,\vec{p}_{2\pi})^2}{\bar{z}s}p_2^{\mu}
     + (-l_\perp +\bar{z}\, p_{2\pi \perp})^{\mu} ,
\end{equation}
where $2\vec{l}$ is the relative transverse momentum of the quarks forming
the two pion system and $\bar{z} = 1-z $, up to small corrections of the
order $\vec{p}_{2\pi}^{\;2}/s$. 
Following the collinear approximation of the
factorization procedure in the description of the two pion formation through the
generalized distribution amplitude, we put $\vec{l}=\vec{0}$ in the hard amplitude.

In a similar way as in (\ref{l1}), (\ref{l2}) we parametrize the momenta of
the produced pions as
\begin{equation}  \label{p+}
p_+^{\mu} = \zeta p_1^{\mu} 
  + \frac{m_\pi^2+(\vec{p}+ \zeta\,\vec{p}_{2\pi})^2}{\zeta s}p_2^{\mu}
  + (p_\perp +\zeta \,p_{2\pi \perp})^{\mu} ,
\end{equation}
\begin{equation}  \label{p-}
p_-^{\mu} = \bar{\zeta} p_1^{\mu} 
  + \frac{m_\pi^2+(-\vec{p}+ \bar{\zeta }\,\vec{p}_{2\pi})^2}{\bar{\zeta }s}p_2^{\mu}
  + (-p_\perp +\bar{\zeta }\, p_{2\pi\perp})^{\mu} ,
\end{equation}
where $2\vec{p}$ is now their relative transverse momentum, 
$\zeta = \frac{p_2\cdot p_+}{p_2\cdot p_{2\pi}}$ is the fraction of
the longitudinal momentum $p_{2\pi}$ carried by the produced $\pi^+$, and 
$\bar{\zeta } = 1-\zeta $.  The variable $\zeta$ is related to the polar
 angle $\theta$ which is defined in the rest frame of the pion pair by
\begin{equation}  \label{theta}
\beta \cos \theta = 2\zeta -1\,,\;\;\; \beta \equiv \sqrt{1 - \frac{%
4\,m_\pi^2}{m_{2\pi}^2}}\;.
\end{equation}
Since the ``longitudinal part'' of the two pion wave function depends only on
the angle $\theta$ and does not depend on the azimuthal decay angle $\phi$
(in the same rest frame of the pair), we focus on the calculation of
charge asymmetries expressed in terms of $\theta$ (see below).
Similar expressions as Eqs.~(\ref{l1}-\ref{theta}) are used for the lower quark loop.
Since we are interested in photon interactions in ultraperipheral collisions of hadrons, the photons are quasi real and hence predominantly transversely polarized. 
The polarization vectors of the photons are written as :
\begin{equation}
\vec{\varepsilon}(+)=-\frac{1}{\sqrt{2}}(1,i),\quad \vec{\varepsilon}(-)=\frac{1}{\sqrt{2}}(1,-i).
\end{equation} 
We will consider spin averaged cross sections since  
hadron colliders do not produce polarized photon beams.

\section{Scattering amplitudes}

It is well known that at high energies ($s\gg |t|$) the amplitude factorizes into impact factors convoluted over the two-dimensional transverse momenta of the $t$-channel gluons.

For the Pomeron exchange, which corresponds in the Born approximation of
QCD to the
exchange of two gluons in a color singlet state, see Fig.~\ref{fig:1},  the impact
representation
has the form
\begin{equation}  \label{pom}
\mathcal{M}_\pom = -i\,s\,\int\;\frac{d^2 \vec{k}_1 \; d^2 \vec{k}_2 \;
\delta^{(2)}(\vec{k}_1 +\vec{k}_2-\vec{p}_{2\pi})}{(2\pi)^2\,\vec{k}_1^2\,%
\vec{k}_2^2} J_\pom^{\gamma_T}(\vec{k}_1,\vec{k}_2)\cdot
J_\pom^{\gamma_T}(\vec{k}_1,\vec{k}_2) ,
\end{equation}
where $J_\pom^{\gamma_T}(\vec{k}_1,\vec{k}_2)$  is the impact factor
for the transition $\gamma_T \to \pi^+\ \pi^-$ via Pomeron exchange.

The corresponding representation for the Odderon exchange, \emph{%
i.e.} the exchange of three gluons in a color singlet state, is given by
the formula
\begin{equation}  \label{odd}
\mathcal{M}_\odd =-\frac{8\,\pi^2\,s}{3!}\int\;\frac{d^2 \vec{k}_1 \; d^2 \vec{k%
}_2\; d^2 \vec{k}_3\; \delta^{(2)}(\vec{k}_1 +\vec{k}_2 +\vec{k}_3-\vec{p}%
_{2\pi})}{(2\pi)^6\,\vec{k}_1^2\,\vec{k}_2^2\,\vec{k}_3^2} 
J_\odd^{\gamma_T}(\vec{k}_1,\vec{k}_2,\vec{k}_3)\cdot 
J_\odd^{\gamma_T}(\vec{k}_1,\vec{k}_2,\vec{k}_3) ,
\end{equation}
where $J_\odd^{\gamma_T}(\vec{k}_1,\vec{k}_2,\vec{k}_3)$ is the
impact factor  for the transition $\gamma_T \to \pi^+\ \pi^-$ via Odderon exchange.

The  impact factors are calculated by the use of standard methods, see e.g.
Ref.~\cite{GI} and references therein.

\subsection{Impact factors for $\gamma \rightarrow \pi ^{+}\pi ^{-} $}

The leading order calculation in pQCD of the  impact factors gives 
\begin{equation}
J_\pom^{\gamma _{T}}(\vec{k}_{1},\vec{k}_{2})=-\frac{i\,e\,g^{2}\,\delta
^{ab}}{4\,N_{C}}\,\int\limits_{0}^{1}\,dz\,(z-{\bar{z}})\;\vec{\varepsilon}%
(T)\cdot \vec{Q}_\pom(\vec{k}_{1},\vec{k}_{2})\;\Phi ^{I=1}(z,\zeta ,m_{2\pi
}^{2}) , \label{IFPT}
\end{equation}
where the vector $\vec{Q}_\pom(\vec{k}_{1},\vec{k}_{2})$ is defined by
\begin{eqnarray}
\label{QP}
\vec{Q}_\pom(\vec{k}_{1},\vec{k}_{2})= 
   \frac{z\,\vec{p}_{2\pi }}{z^{2}\vec{p}_{2\pi }^{\;2}+\mu ^{2}}
 - \frac{{\bar{z}}\,\vec{p}_{2\pi }}{{\bar{z}}^{2}\vec{p}_{2\pi }^{\;2}+\mu ^{2}}
 + \frac{\vec{k}_{1}-z\,\vec{p}_{2\pi }}{(\vec{k}_{1}-z\,\vec{p}_{2\pi })^{2}+\mu ^{2}}
 - \frac{\vec{k}_{1}-{\bar{z}}\,\vec{p}_{2\pi }}{(k_{1}-{\bar{z}}\,\vec{p}_{2\pi})^{2}+\mu ^{2}}  .
\end{eqnarray}
The calculation of the Odderon exchange contribution gives 
\begin{equation}
J_\odd^{\gamma _{T}}(\vec{k}_{1},\vec{k}_{2},\vec{k}_{3})=
-\frac{i\,e\,g^{3}\,d^{abc}}{8\,N_{C}}\,
\int\limits_{0}^{1}\,dz\,(z-{\bar{z}})\;\vec{\varepsilon}(T)\cdot 
\vec{Q}_\odd(\vec{k}_{1},\vec{k}_{2},\vec{k}_{3})\;
\frac{1}{3}\Phi ^{I=0}(z,\zeta ,m_{2\pi }^{2}) , \label{IFOT}
\end{equation}
where we have used the definition
\begin{eqnarray}
\label{QO}
\vec{Q}_\odd(\vec{k}_{1},\vec{k}_{2},\vec{k}_{3})
&=&\frac{z\,\vec{p}_{2\pi }}{z^{2}\vec{p}_{2\pi }^{\;2}+\mu ^{2}}
  + \frac{{\bar{z}}\,\vec{p}_{2\pi }}{{\bar{z}}^{2}\vec{p}_{2\pi }^{\;2}+\mu ^{2}}
 + \sum\limits_{i=1}^{3}
    \left( \frac{\vec{k}_{i}-z\,\vec{p}_{2\pi }}{(\vec{k}_{i}-z\,\vec{p}_{2\pi })^{2}+\mu ^{2}}
          +\frac{\vec{k}_{i}-{\bar{z}}\,\vec{p}_{2\pi }}{(\vec{k}_{i}-{\bar{z}}\,\vec{p}_{2\pi })^{2}+\mu ^{2}}\right) .  
\end{eqnarray}

\subsection{Generalized two pion distribution amplitudes}

A crucial point of the present study is the choice of an appropriate
two pion distribution amplitude (GDA) \cite{Hagler:2002nh,Hagler:2002sg,Hagler:2002nf,Hagler:2002cq,Diehl:1998dk,Diehl:2000uv,Ivanov:2006wm,Warkentin:2007su} which includes
the full strong interaction related to the production of the two pion system.

Based on  an expansion of the GDA in Gegenbauer polynomials $C_n^m(2z-1)$ and in Legendre polynomials $P_l(2\zeta-1)$ \cite{Polyakov:1998ze,Kivel:1999sd} it is believed that only the first terms give a significant contribution:
\begin{eqnarray}
  \Phi^{I=1} (z,\zeta,m_{2\pi}) &=& 6z\zbar\beta f_1(m_{2\pi}) \cos\theta ,\\
  \Phi^{I=0} (z,\zeta,m_{2\pi}) &=& 5z\zbar(z-\zbar)\left[-\frac{3-\beta^2}{2}f_0(m_{2\pi})+\beta^2f_2(m_{2\pi})P_2(\cos\theta)\right],
\end{eqnarray}
where $f_1(m_{2\pi})$ can be identified with the electromagnetic pion form factor $F_\pi(m_{2\pi})$. For our calculation we use the following $F_\pi$-parametrization  inspired by Ref.~\cite{Kuhn:1990ad}
\begin{equation}
f_1(m_{2\pi})=  F_\pi(m_{2\pi}) = \frac{1}{1+b}BW_\rho(m_{2\pi}^2)\frac{1+a BW_\omega(m_{2\pi}^2)}{1+a},
\end{equation}
with
\begin{eqnarray}
  BW_\rho(m_{2\pi}^2)     &=& \frac{m_\rho^2}{m_\rho^2-m_{2\pi}^2-i\,m_{2\pi}\Gamma_\rho(m_{2\pi}^2)} \\
  \Gamma_\rho(m_{2\pi}^2) &=& \Gamma_\rho\frac{m_\rho^2}{m_{2\pi}^2}\left(\frac{m_{2\pi}^2-4m_{\pi}^2}{m_\rho^2-4m_{\pi}^2}\right)^{3/2} \\
  BW_\omega(m_{2\pi}^2)   &=& \frac{m_\omega^2}{m_\omega^2-m_{2\pi}^2-i\,m_\omega\Gamma_\omega}.
\end{eqnarray}
As masses and widths we use $m_\rho=775.49\,{\rm MeV}$, $\Gamma_\rho=146.2\,{\rm MeV}$, $m_\omega=782.65\,{\rm MeV}$, $\Gamma_\omega=8.49\,{\rm MeV}$ \cite{PDG}. We fit the remaining free parameters  to the data compiled in Ref.~\cite{Barkov:1985ac} obtaining $a=1.78\cdot 10^{-3}$ and $b=-0.154$. 
Including a hypothetical $\rho'$ resonance as originally used in Ref.~\cite{Kuhn:1990ad}, gives a significant better fit to the data at large $m_{2\pi}$ but has only small effect on the asymmetry which is the main object of our studies.

 For the $I=0$ component we use different models. The first model follows Ref.~\cite{Hagler:2002nh} and reads
\begin{equation}
  f_{0/2}(m_{2\pi}) = e^{i\delta_{0/2}(m_{2\pi})}\left| BW_{f_{0/2}}(m_{2\pi}^2)\right| .
\end{equation}
The phase shifts $\delta_{0/2}$ are those from the elastic $\pi^+\pi^-$ scattering, for which we use the parametrization of Ref.~\cite{Pelaez:2004vs} below  and that of Ref.~\cite{Kaminski:2006yv} above the $K\bar{K}$ threshold.

$| BW_{f_{0/2}}(m_{2\pi}^2)|$ is the modulus of the Breit-Wigner amplitudes
\begin{equation}
  BW_{f_{0/2}}(m_{2\pi}^2) = \frac{m_{f_{0/2}}^2}{m_{f_{0/2}}^2-m_{2\pi}^2-i\,m_{f_{0/2}}\Gamma_{f_{0/2}}} ,
\end{equation}
with $m_{f_{0}}=980\,{\rm MeV}$, $\Gamma_{f_{0}}=40-100\,{\rm MeV}$, $m_{f_{2}}=1275.1\,{\rm MeV}$, $\Gamma_{f_{2}}=185\,{\rm MeV}$ \cite{PDG}.

In the second model -- elaborated in Ref.~\cite{Warkentin:2007su} -- the functions $f_{0/2}$ are the corresponding Omn\`es functions for $S-$ and $D-$waves constructed by dispersion relations from the phase shifts of the elastic pion scattering:
\begin{equation}
  f_l(m_{2\pi}) = \exp\left(\pi I_l + \frac{m_{2\pi}^2}{\pi}\int_{4m_\pi^2}^{\infty}ds\frac{\delta_l(s)}{s^2(s-m_{2\pi}^2-i \varepsilon)}\right),\quad {\rm with}\quad
  I_l = \frac{1}{\pi}\int_{4m_\pi^2}^{\infty}ds\frac{\delta_l(s)}{s^2}.
\end{equation}

The assumption that the phases of the GDA equal those of the elastic scattering looses its solid base beyond the $K\bar{K}$ threshold. As discussed in Refs.~\cite{Warkentin:2007su,Ananthanarayan:2004xy} it might well be that the actual phases of the GDA are closer to the phases of the corresponding $\mathcal{T}$ matrix elements $\frac{\eta_l e^{2i\delta_l}-1}{2i}$. The third model for the $I=0$ component of the GDA takes this into account by using the technique of model 2 with these phases $\delta_{\mathcal{T},l}$ of the $\mathcal{T}$ matrix elements. 

While the first and the second model give quite compatible results, model three differs from them significantly. The most striking difference is the absence of pronounced $f_0$ resonance effects in model 3. In fact, measurements at HERMES \cite{Airapetian:2004sy} do not observe a resonance effect at the $f_0$-mass even though a confirmation by an independent experiment would be desirable. From the same measurements at HERMES one can draw the conclusion that using $\delta_{T,2}$ and $\delta_2$ for the $f_2$ region are both compatible with data \cite{Warkentin:2007su}. Having this in mind, we consider also a fourth model -- a mixed description with the $f_0$ contribution from model 3 and the $f_2$ contribution from model 2.

\subsection{Photon exchange amplitude}

The photon has the same $C$-parity as the Odderon. Therefore, its exchange between the two quark-antiquark systems can mimic an Odderon exchange. The according amplitude is straightforward to calculate and reads
\begin{equation}  \label{gam}
\mathcal{M}_\gamma =\frac{s}{2t} J_\gamma^{\gamma_T}\cdot J_\gamma^{\gamma_T},
\end{equation}
with
\begin{equation}
J_{\gamma}^{\gamma _{T}}=
\frac{e^2}{2}\,\int\limits_{0}^{1}\,dz\,(z-{\bar{z}})\;
\vec{\varepsilon}(T)\cdot \vec{p}_{2\pi}
\left(\frac{z}{\mu^2+z^2\vec{p}_{2\pi}^{\;2}}+\frac{\zbar}{\mu^2+\zbar^2\vec{p}_{2\pi}^{\;2}}\right)
\Phi ^{I=0}(z,\zeta ,m_{2\pi }^{2})  \label{IFGT}.
\end{equation}

In our concrete process the photon exchange amplitude amounts to the order of 10\% of the Odderon exchange amplitude. Although we do not neglect this contribution, it is clear that the asymmetry described in the following section is driven by the Odderon/ Pomeron-interference.

\section{Charge asymmetries and rates}

In the following we make use of the fact that the GDAs for $C$-even and for $C$-odd pion pairs are orthogonal to each other in the space of Legendre polynomials in $\cos\theta$. As a consequence, in the total cross section the interference term completely vanishes. In contrast only the interference term survives, when the amplitude squared is multiplied by $\cos\theta$ before the angular integration which corresponds to selecting the charge asymmetric contribution.
The asymmetry we are interested in is defined as
\begin{eqnarray}
  A(t,m_{2\pi}^2,m_{2\pi}'^2) &=& \frac{\int\cos\theta\,\cos\theta'\,d\sigma(t,m_{2\pi}^2,m_{2\pi}'^2,\theta,\theta')}{\int\,d\sigma(t,m_{2\pi}^2,m_{2\pi}'^2,\theta,\theta')} \nonumber\\
 &=& \frac{\int_{-1}^1d\cos\theta\int_{-1}^1d\cos\theta'\;2\cos\theta\,\cos\theta'\,{\rm Re}\left[\mathcal{M}_\pom(\mathcal{M}_\odd+\mathcal{M}_{\gamma})^*\right]}{\int_{-1}^1d\cos\theta\int_{-1}^1d\cos\theta'\,\left[\left|\mathcal{M}_\pom\right|^2+\left|\mathcal{M}_\odd+\mathcal{M}_{\gamma}\right|^2\right]} .
\end{eqnarray}

The obtained landscape as a function of the two invariant masses is not particularly illuminating and would be difficult to measure.
To reduce the complexity, we integrate over the invariant mass of one of the two pion systems to obtain
\begin{eqnarray}
  \hat{A}(t,m_{2\pi}^2;m_{\rm min}^2,m_{\rm max}^2) &=& \frac{\int_{m_{\rm min}^2}^{m_{\rm max}^2} dm_{2\pi}'^2\int\cos\theta\,\cos\theta'\,d\sigma(t,m_{2\pi}^2,m_{2\pi}'^2,\theta,\theta')}{\int_{m_{\rm min}^2}^{m_{\rm max}^2} dm_{2\pi}'^2\int\,d\sigma(t,m_{2\pi}^2,m_{2\pi}'^2,\theta,\theta')} \nonumber\\
 &=& \frac{\int_{m_{\rm min}^2}^{m_{\rm max}^2} dm_{2\pi}'^2\int_{-1}^1d\cos\theta\int_{-1}^1d\cos\theta'\;2\cos\theta\,\cos\theta'\,{\rm Re}\left[\mathcal{M}_\pom(\mathcal{M}_\odd+\mathcal{M}_{\gamma})^*\right]}{\int_{m_{\rm min}^2}^{m_{\rm max}^2} dm_{2\pi}'^2\int_{-1}^1d\cos\theta\int_{-1}^1d\cos\theta'\,\left[\left|\mathcal{M}_\pom\right|^2+\left|\mathcal{M}_\odd+\mathcal{M}_{\gamma}\right|^2\right]}. \label{eq:ahat}
\end{eqnarray}

Let us note that since two pion pairs are always in the same $C$-parity 
state, because of $C=+$ parity of the initial $\gamma \gamma$ state, it is 
necessary to keep in Eq.~\eqref{eq:ahat} the integration weight  $\cos \theta'$.
Without this weight the single charge asymmetry would vanish. The deviations from 
this vanishing of the asymmetry may serve as a 
measure of experimental uncertainties.

\begin{figure}
  \centering
  \includegraphics[width=8cm]{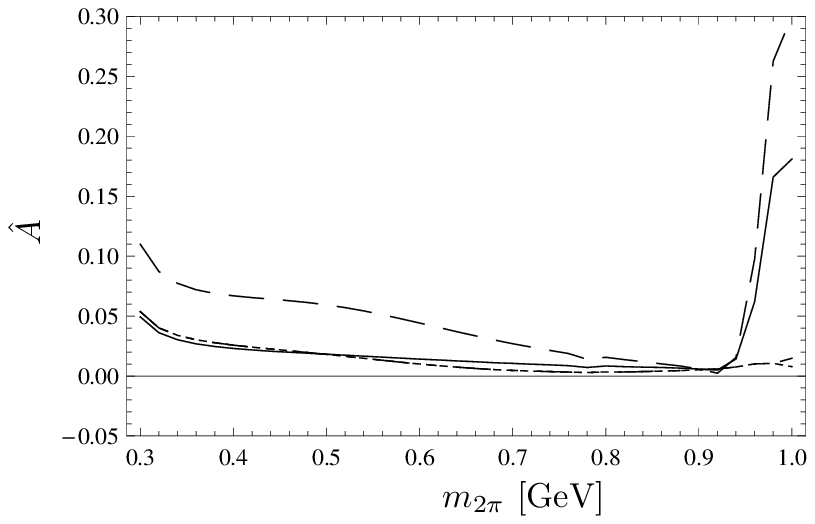}\hspace{10mm}
  \includegraphics[width=8cm]{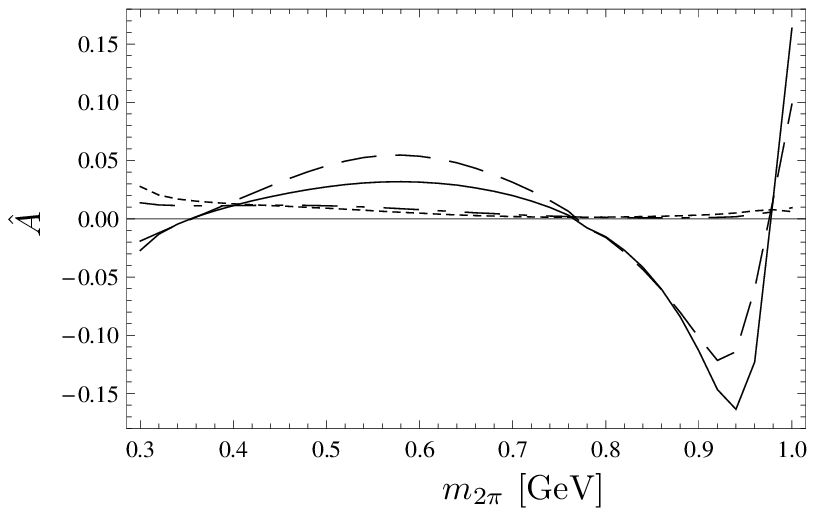}\\
  \caption{Asymmetry $\hat{A}$ at $t=-1\,\gev^2$ for model 1 (solid), 2 (dashed), 3 (dotted), and 4 (dash-dotted) -- model 3 and 4 are nearly on top of each other. Left column has $m_{\rm min}=.3\,\gev$ and $m_{\rm max}=m_\rho$, while right column has $m_{\rm min}=m_\rho$ and $m_{\rm max}=1\,\gev$. 
}
\label{fig:asymplot1}
\end{figure}

\begin{figure}
  \centering
  \includegraphics[width=8cm]{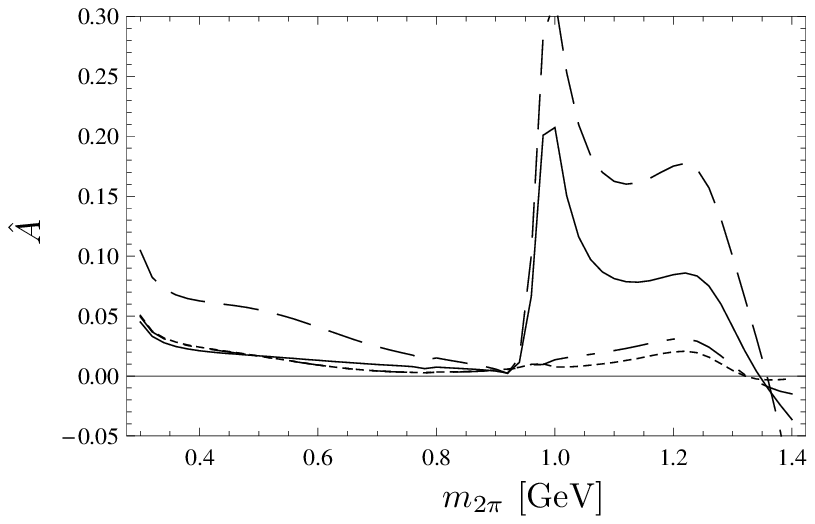}\hspace{10mm}
  \includegraphics[width=8cm]{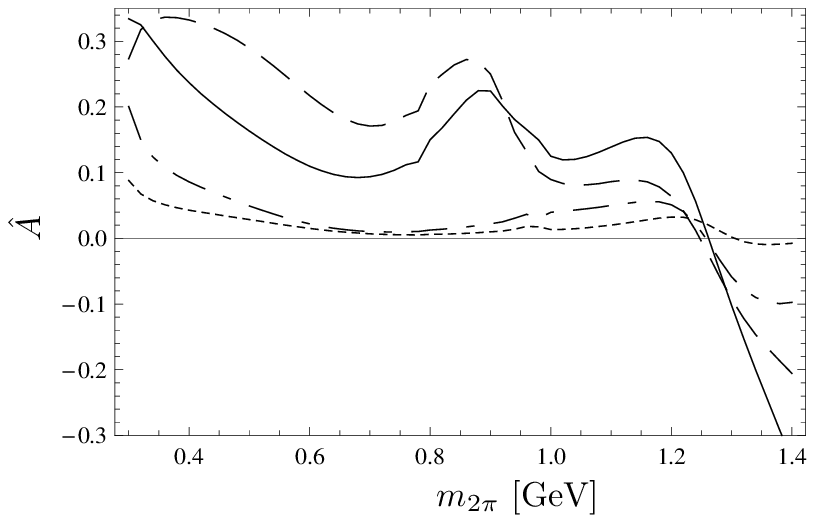}\\
  \caption{Asymmetry $\hat{A}$ at $t=-2\,\gev^2$ for model 1 (solid), 2 (dashed), 3 (dotted), and 4 (dash-dotted). Left column has $m_{\rm min}=.3\,\gev$ and $m_{\rm max}=m_\rho$, while right column has $m_{\rm min}=m_{f_0}$ and $m_{\rm max}=1.4\,\gev$. 
}
\label{fig:asymplot2}
\end{figure}

Considering an analytic calculation of the matrix element in Eq.~(\ref{odd}), it turns out that it would require the calculation of two dimensional two-loop box diagrams, whose off-shellness for all external legs is different. 
The techniques developed in Refs.~\cite{Broadhurst:1993ib,Pire:2005ic,Segond:2007fj} can not be applied here due to the very elaborated topology of the most complicated diagrams involved (square box with one additional diagonal line).
Instead we rely on a numerical evaluation by Monte Carlo methods. In particular we make use of a modified  version of {\sc Vegas} as it is provided by the {\sc Cuba} library \cite{Hahn:2004fe}.
The result for the asymmetry $\hat{A}$ at $t=-1\,\gev^2$ (resp. $t=-2\,\gev^2$)  is shown in Fig.~\ref{fig:asymplot1} (resp. Fig.~\ref{fig:asymplot2}).
Since our framework is only justified for $m_{2\pi} ^2 < -t$, (in fact strictly speaking, one even needs $m_{2\pi}^2 \ll -t$ ), we keep $m_{2\pi}$ below 1\,GeV  (resp 1.4\,GeV). 
In each case, we present the expected asymmetry with the GDAs parametrized as discussed above.

\begin{figure}
  \centering
  \includegraphics[width=10cm]{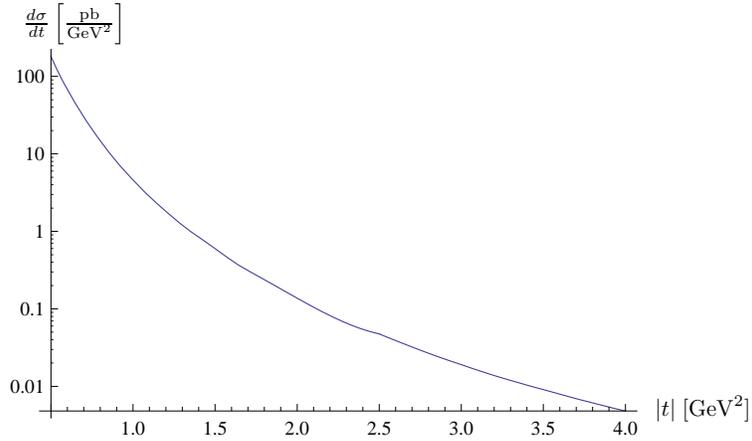}
  \caption{$t$-dependence of photon-photon cross section.}
  \label{fig:dsigmadt}
\end{figure}

In order to evaluate the feasibility of the Odderon search, we need to  supplement the calculation of the asymmetry with rate estimates in ultraperipheral collisions at hadron colliders. This rate depends on the Pomeron dominated photon-photon cross section and on the equivalent photon flux.
The total photon-photon cross section falls off rapidly with increasing $|t|$ (see Fig.~\ref{fig:dsigmadt}). Therefore, the integration mainly depends on the lower limit of $|t|$-integration ($\tmin$). Already for $\tmin=-1\,\gev^2$ we find $\sigma_{\gamma\gamma}=1.1\,{\rm pb}$. 

Although in  Ref.~\cite{Baur:2001jj} it is claimed that the photon flux is best for medium-weight ions, and especially superior to that of protons, this is in fact not true. To obtain Fig.~6 of Ref.~\cite{Baur:2001jj} the effective $\gamma\gamma$ luminosity has been calculated for protons and ions by the Monte Carlo program {\sc Tphic} \cite{Hencken:1996ge} which is based on the Weizs{\"a}cker-Williams method \cite{vonWeizsacker:1934sx,Williams:1934ad} with the additional condition of non-overlapping ions \cite{Baur:1990fx,Cahn:1990jk}, but the authors used a quite small luminosity for proton-proton collisions at the LHC ($14\,000\,{\rm mb}^{-1}{\rm s}^{-1}$ instead of the official design luminosity $10^{7}\,{\rm mb}^{-1}{\rm s}^{-1}$ \cite{Bosser:2000kb,Bruning:2004ej}). Very unfortunately, the identical figure is reprinted in Ref.~\cite{Baltz:2007kq} while a non consistent p-p luminosity of $10^{7}\,{\rm mb}^{-1}{\rm s}^{-1}$ is cited.

As already was shown by Cahn and Jackson \cite{Cahn:1990jk}, the $\gamma\gamma$ luminosity in case of ions can be expressed in terms of a universal function $\xi(z)$, where $z=M R\approx 5.665 A^{4/3}$ with $M$ being the mass and $R$ the radius of the ion, and a prefactor proportional to $Z^4$. Since the luminosity for ions at the LHC decreases roughly as $Z^{-4}$, the prefactor's $Z$ dependence is more or less compensated and only the universal function $\xi$ remains which is exponentially decreasing with $z$. Hence, lighter projectiles provide a larger effective $\gamma\gamma$ luminosity, with the protons offering the highest luminosity. 

\begin{figure}
  \centering
   \includegraphics[width=14cm]{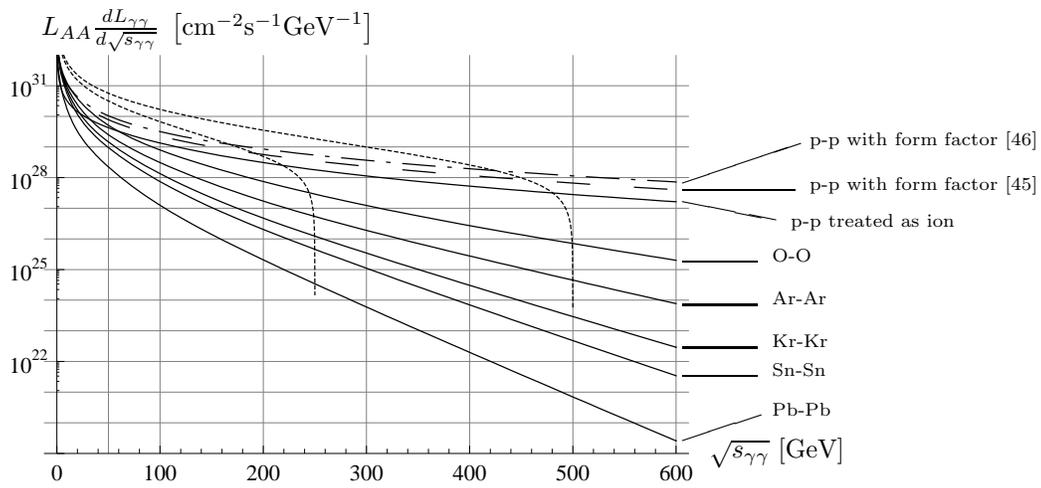}
  \caption{Effective $\gamma\gamma$ luminosities for the collision of p-p based on Ref.~\cite{Drees:1988pp} (dash-dotted) and Ref.~\cite{Nystrand:2004vn} (dashed). The results using the parametrization of Ref.~\cite{Cahn:1990jk} for ions are given by solid lines for p-p,
${\rm O}^{8}_{16}$-${\rm O}^{8}_{16}$,
${\rm Ar}^{18}_{40}$-${\rm Ar}^{18}_{40}$,
${\rm Kr}^{36}_{84}$-${\rm Kr}^{36}_{84}$,
${\rm Sn}^{50}_{120}$-${\rm Sn}^{50}_{120}$,
${\rm Pb}^{82}_{208}$-${\rm Pb}^{82}_{208}$ from top to bottom. For ions we used the average luminosities as given in Ref.~\cite{Brandt:2000mu}, for proton we used $L_{pp}=10^{34}\,{\rm cm}^{-2}{\rm s}^{-1}$. For comparison also effective $\gamma\gamma$ luminosities at the ILC are given for $\sqrt{s_{e^+e^-}}=250\,{\rm GeV}$ and $\sqrt{s_{e^+e^-}}=500\,{\rm GeV}$ (both as dotted lines).}
  \label{fig:lumi}
\end{figure}

Therefore, we provide a corrected overview over the various effective $\gamma\gamma$ luminosities in Fig.~\ref{fig:lumi}. For the different ion scenarios that are discussed in Ref.~\cite{Brandt:2000mu} we use the parametrization of Ref.~\cite{Cahn:1990jk} which relies on the Weizs{\"a}cker-Williams method \cite{vonWeizsacker:1934sx,Williams:1934ad} with the additional condition of non-overlapping ions \cite{Baur:1990fx,Cahn:1990jk}. 
For protons usually a calculation based on the proton electric dipole form factor  $F_E(Q^2) = 1/(1+\frac{Q^2}{0.71\,\gev^2})$ in combination with the Weizs{\"a}cker-Williams method is used, as it can be found in Ref.~\cite{Drees:1988pp}.
In Ref.~\cite{Nystrand:2004vn} a slightly improved version is given, which lowers the photon flux. An inclusion of the corresponding magnetic dipole moment \cite{Kniehl:1990iv} would lead to a flux between those both. 
For this reason, we use the formulas given in Ref.~\cite{Drees:1988pp} and Ref.~\cite{Nystrand:2004vn} for the case of proton-proton collisions. We also provide the results for the proton treated as a heavy ion because it might be that the non-overlap condition  -- reducing the flux -- is of importance \cite{Nystrand:2004vn}, even if such a procedure does not include the proton form factor. We consider such a lower result as a conservative lower estimate.
Since the luminosity factors will cancel in the asymmetry, this uncertainty does not affect our conclusions on the Odderon effects.
For comparison, we provide also the effective $\gamma\gamma$ luminosities at the intended ILC where the design luminosity for $e^+e^-$ collisions is $2\cdot 10^{34}\,{\rm cm}^{-2}{\rm s}^{-1}$ \cite{Brau:2007zza} \footnote{In the calculation of the equivalent photon spectrum \cite{Budnev:1974de} we used $q_{\rm max}=100\,{\rm GeV}$.}.

\begin{figure}
  \centering
  \includegraphics[width=8cm]{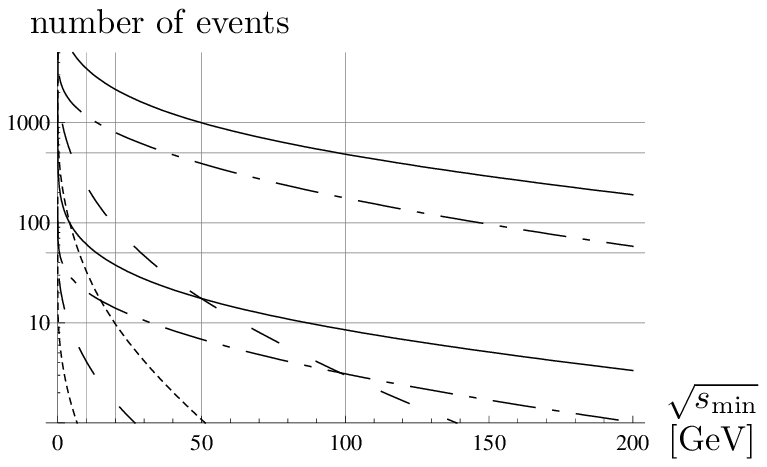}\hspace{10mm}\includegraphics[width=8cm]{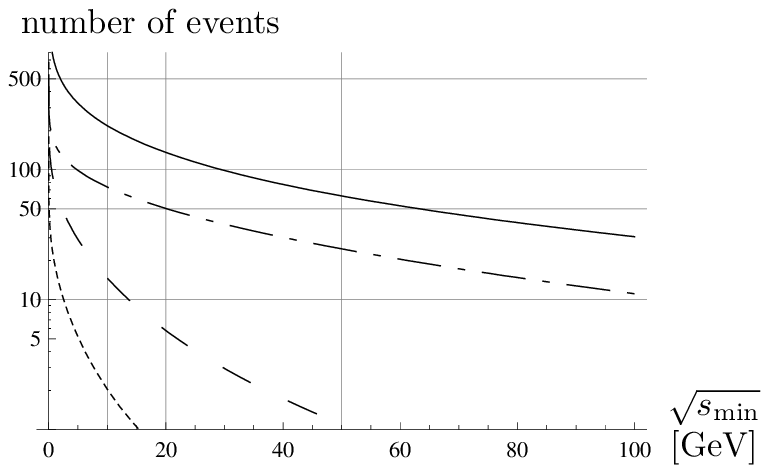}
  \caption{Rate of production of two pion pairs in ultraperipheral collisions in dependence on the lower cut $\smin$ given in `events per month' in case of ions, and `events per six months' in case of protons which in both cases correspond to one year of running of LHC. The l.h.s. plot shows the rate for $\tmin=-1\,\gev^2$, the r.h.s. that for  for $\tmin=-2\,\gev^2$. The solid line displays the result for p-p collision using Ref.~\cite{Drees:1988pp}, the dashed-dotted that for protons treated as heavy ions, the dashed one that for Ar-Ar collisions, and the dotted line that for Pb-Pb collisions.
On the left figure, also the much smaller rates coming from the Odderon exchange are shown (with the same dashing).}
\label{fig:rates}
\end{figure}

As shown in Fig.~\ref{fig:lumi}, the effective $\gamma\gamma$ luminosity decreases rapidly with increasing energy.
Since our intermediate hard scale $t$ is quite small, we are not forced to consider extremely large photon-photon energies. The only condition on the minimal photon-photon energy ($\smin$) is
$\smin\gg |t]$ to ensure the validity of high energy factorization into two separated pion systems. 
Hence, a $\smin$ of $400\,\gev^2$ could be already enough. The effect of varying $\smin$ is displayed in Fig.~\ref{fig:rates}. 

For p-p collisions, the rates are high and even for $\tmin=-2\,\gev^2$ sizable. Although heavy ions would offer the possibility to trigger on ultraperipheral collisions by detecting neutrons from giant dipole resonances (GDR) in the Zero Degree Calorimeters, the rates that can be read off from Fig.~\ref{fig:rates} are rather low. Only for medium-weight ions there might be the possibility to measure the process.

In hadron-hadron collisions the process of interest could as well be connected by Pomerons to the colliding hadrons. In that case one would have to deal with the unknown hadron-Pomeron and Pomeron-Pomeron-two-pion couplings. Therefore, we consider that process as background. This circumstance would make heavy ions preferable because of the different scaling of Pomeron and photon coupling when changing from proton to nuclei scattering. However, experimentally such a background can be suppressed by refusing events with a total $p_T$ larger than some small cut-off.
Indeed, the $\gamma\gamma$ events dominates, due to the photon propagator 
singularities, when each of the transverse momentum is small, which on the 
average is satisfied when the {\it total} transverse momentum is imposed 
to be small.

Inclusion of high energy evolution of the two gluon exchange \`a la BFKL \cite{Fadin:1975cb,Kuraev:1976ge,Kuraev:1977fs,Balitsky:1978ic,Fadin:1998py,Ciafaloni:1998gs} would increase the total cross section but at the same time diminish the asymmetry since the Odderon including  BKP evolution \cite{Bartels:1980pe,Kwiecinski:1980wb} has a  smaller intercept. 
The study of these effects goes beyond the goal of this paper. Since we demonstrated that the rate and asymmetry were sizable for moderate values of the $\gamma \gamma$ energy, we do not expect evolution effects to play a major role in the proposed strategy to uncover Odderon effects.

\section{Conclusion}

We have investigated in real photon-photon collision the production of two $\pi^+\pi^-$-pairs well separated in rapidity. Due to the non fixed $C$-parity of these pairs, beside a Pomeron exchange an Odderon exchange is possible as well. We have calculated both contributions in a perturbative approach -- justified by $t$ providing the hard scale -- where the only soft building blocks needed are the GDAs of the pion pairs. 
We have shown that a charge asymmetry in the polar angle $\theta$ (defined in the rest frame of the pion pairs) is linearly dependent on the Odderon amplitude and moreover is sizable but GDA-model dependent.
In fact, the predicted asymmetry depends much of the two pion GDAs, which are  
still not really known although HERMES measurements of two pion  
electroproduction \cite{Airapetian:2004sy}  disfavor models with a strong $f_0$ coupling to  
the $\pi^+ \pi^-$ state. We however think that higher statistics data, which  
may come from a JLab experiment at 6 or 12\,GeV, are definitely needed  
before one can trust a definite model of the GDAs. Because two pion  
deep electroproduction in the low energy domain is dominated by quark  
exchanges, this test of the GDA models is independent of any Odderon  
search. This looks like a prerequisite to a trustable extraction of  
the Odderon signal -- or of an upper bound on Odderon exchange  
amplitudes -- from ultraperipheral collisions.

We have discussed the possibility to measure this process in ultraperipheral collisions at LHC. 
Although in ion-ion collision it is easier to trigger on ultraperipheral collision with a lower background from diffractive events from strong interactions, the event rates are too low. In proton-proton collision the rates are high enough to measure the asymmetry even though isolating ultraperipheral collisions in proton-proton collisions may be a challenge for experimentalists.

Let us finally note that the background from strong interactions would be completely absent in an $e^+e^-$ collider which via Compton-back-scattering could work as a very effective $\gamma\gamma$ collider. At the ILC \cite{Brau:2007zza} for a nominal electron beam energy of 250\,GeV the luminosity for photon-photon collisions would be $3.5\cdot 10^{33}\,{\rm cm}^{-2}{\rm s}^{-1}$ \cite{Telnov:2005ad} for the high energy photons (energy fraction at least 80\% of the maximal possible energy fraction) or even higher, if optimized for photon-photon collisions \cite{Telnov:2006cj}. As Fig.~\ref{fig:lumi} reveals, even running in the electron-positron mode the effective $\gamma\gamma$ luminosity is slightly larger than at the LHC. For these reasons, the ILC would provide an ideal environment to study the process of interest.

{\it Acknowledgments.}  
We acknowledge discussions with Mike Albrow, Gerhard Baur, David d'Enterria, Bruno Espagnon, and Rainer Schicker.
This work is supported in part by the Polish Grant N N202 249235, the French-Polish scientific agreement Polonium, by the grant ANR-06-JCJC-0084 and by the ECO-NET program, contract 12584QK.

\end{document}